\documentclass[12pt,a4paper]{revtex4}
\usepackage{multirow}
\usepackage{amssymb}
\usepackage{amsmath,graphicx}
\usepackage{array}
\usepackage{braket}
\usepackage{bm}
\usepackage{enumerate}
\usepackage{epstopdf}
\usepackage{amsmath}
\usepackage{dcolumn,caption,booktabs}
\usepackage{slashed}
\usepackage{amsfonts}
\usepackage{adjustbox}
\usepackage{tcolorbox}
\usepackage{float}
\usepackage{collectbox}
\usepackage{listings}
\usepackage[unicode]{hyperref}
\usepackage[mathscr]{eucal}
\usepackage{graphics}
\usepackage{subcaption}
\newcommand{\be}{\begin{equation}}
	\newcommand{\ee}{\end{equation}}
\newcommand{\bea}{\begin{eqnarray}}
	\newcommand{\eea}{\end{eqnarray}}
\newcommand{\ben}{\begin{enumerate}}
	\newcommand{\een}{\end{enumerate}}
\newcommand{\bde}{\begin{widetext}}
	\newcommand{\ede}{\end{widetext}}
\newcommand{\nn}{\nonumber}
\newcommand{\crn}{\nonumber \\}

\newcommand{\al}{\alpha}
\newcommand{\la}{\lambda}

\newcommand{\ga}{\gamma}
\newcommand{\va}{\varphi}
\newcommand{\om}{\omega}

\newcommand{\fr}{\frac}

\newcommand{\bc}{\begin{center}}
	\newcommand{\ec}{\end{center}}

\newcommand{\ka}{\kappa}
\newcommand{\La}{\Lambda}
\newcommand{\si}{\sigma}

\newcommand{\Om}{\Omega}

\begin{document}
	
	\title{Electroweak phase transition via Dilaton in Two-Time Physics}
	
	\author{Vo Quoc Phong$^{a,b}$}	
	\email{vqphong@hcmus.edu.vn}
	\affiliation{$^a$Department of Theoretical Physics, University of Science, Ho Chi Minh City 700000, Vietnam\\ $^b$Vietnam National University, Ho Chi Minh City 700000, Vietnam}

	\author{Dam Quang Nam$^{a,b}$}
	\email{dqnam1995@gmail.com}
	\affiliation{$^a$Department of Theoretical Physics, University of Science, Ho Chi Minh City 700000, Vietnam\\ $^b$Vietnam National University, Ho Chi Minh City 700000, Vietnam}

\begin{abstract}
	The Two-time model (2T model) has six dimensions with two dimensions of time, has a Dilaton particle that makes the symmetry breaking differently from the Standard Model. Assuming a soft break of $SP(2,R)$ symmetry,  the 2T extension can give a suitable picture of  the matter-antimatter asymmetry by the Baryogenesis scenario. By reducing the 2T metric to the Minkowski metric (1T metric) and using a new form of Dilaton potential, we consider the electroweak phase transition picture in the 2T model with the Dilaton as a trigger. Our analysis shows that Electroweak Phase Transition (EWPT) is a first-order phase transition at the $200$ GeV scale, its strength is about $1 - 3.08$ and the mass of Dilaton  is in the interval $[345,625]$ GeV. Therefore, the 2T-model indirectly suggests that extra-dimension can  also  be a source of EWPT.
\end{abstract}
\maketitle

\section{INTRODUCTION}\label{secInt}

Since the Higgs boson was discovered at the LHC, Particle Physics almost completed its mission that provides a more accurate understanding of mass. Particle physics is currently in the process of tackling the most important issues of the Standard Model (SM) such as dark matter (DM), dark energy, baryon asymmetry, etc.

In another aspect of Particle Physics, High energy Astrohysics has also been studied from the view of extra dimensions or the viscous  universe model to explain physical-mathematical problems such as hierarchical or cosmic acceleration and inflation. More importantly, these problems have been studied by using models that are connected to high-energy-scale particle physics. Accordingly, the cause of these phenomena may lead to the existence of new particles.

Along with current advances in experimentation, when experiments reach higher energies (a few TeV), it means that one can observe phenomena in the early universe. According to that trend, Cosmology and Particle Physics need to be connected. One of the current research trends is to examine particle theory in higher dimensions. This is an attempt to find the possible combinations of matter and micro-behavior of the matter. String theory is a broad representation of this research problem. However, string theory is difficult to combine with SM. Multidimensional models in cosmology are also not easy to incorporate with SM because they provide only a space-time background but cannot consider particles. Two-time (2T) physics \cite{bars1999,bars2000a,bars2001,survey,bars2006,kuo2006,33} is a suitable model in this direction of research. 2T physics describes particles and their interactions over a space-time larger than 4. This model fully describes the SM particles and proposes a new particle called Dilaton.

For more than a decade (since 2008), the 2T model is a trendy model, is an extended theory of the standard model combined with the theory of space-time or gravity. This model gives us a strange view of time which advances on a plane; that is, it has two dimensions of time. The spontaneous symmetry breaking in this model is also different from that in the SM. More importantly, Dilaton is the new material that we will focus on in this article.

As mentioned, the matter-antimatter asymmetry (baryon asymmetry) is an important topic in Particle Physics and Cosmology. To explain this, there are currently two scenarios: Leptongenesis and Baryogenesis. Baryogenesis requires a more powerful CP-violation source in the SM, new triggers for the first-order electroweak phase transition. According to current research, the triggers can be DM or new heavy bosons  Refs.~\cite{5percent,sakharov,BSM1,BSM2,BSM3,majorana,thdm1,thdm2,ESMCO,phonglongvan1,phonglongvan2,phonglongvan3,phonglongvan4,dssm,munusm,lr,ppf1,ppf2,ppf3,r331a,r331b,r331c,r331d,r331e,Buras:2012dp,1101.4665a,1101.4665b}.

We also recognize that the matter-antimatter asymmetry has a deep connection with the number of dimensions of space-time, especially the number of time dimension increases, the number of particles is likely to increase as well. The 2T model, therefore, has both an extra-dimensional effect and one exotic particle (Dilaton) but this problem is not yet analyzed in detail, i.e, extra dimensions and Dilaton boson can be new materials for the EWPT. Therefore, we will analyze the role of Dilaton in the EWPT problem in this article.

This article is organized as follows. In section \ref{sec2}, we summarize the 2T model and 2T SM with the boson, lepton, and Higgs sectors. In section \ref{sec3}, we derive the effective potential in the 1T model by using the gauge fixing technique, this potential has the contribution from the Dilaton, which is a function of temperature, VEV, and mass of the Dilaton. We analyze in detail the structure of phase transition, find the first-order phase transition, and show constraints on the mass of the Dilaton. Finally, we summarize and make outlooks in section \ref{sec5}.

\section{Standard Model in 2T model}\label{sec2}

Two-time physics (2T model) is a spacetime theory in which the physical phenomena are not different from ones in one-time physics, i.e., the usual spacetime formalism, but it can show a new perspective to investigate some phenomena in ordinary one-time physics. The spacetime in the 2T model has two extra dimensions, one spacelike and one timelike. The basis of the 2T model were established by I. Bars \cite{bars2006} from the string theories and then became a distinctive theory of two-time physics. Moreover, the 2T model resolves the strong CP violation problem, has a candidate for DM and it is a good guide for constructing M-theory \cite{bars2001}.

The SM in 2T model, in particular 4+2 dimensions, could be constructed with the scalar fields such as Higgs $H$ and Dilaton $\Phi$, fermion fields including left/right-handed chiral field $\Psi^L , \Psi^R$ describing the quarks and the leptons, and gauge bosons $A_M$ of the gauge group of the theory \cite{bars2006}.

$M, N$ runs over $4+2$ spacetime dimensional indices which are denoted by $0',0,1',1,2,3$, the notation with apostrophe
refers to the extra dimension, 0-notation denotes the timelike dimension and others are for spatial dimensions as usual.
In the flat spacetime, $\eta_{MN} = \textrm{diag}(-1,-1,1,1,1,1)$.

The internal Yang-Mills group structure is the same as the usual SM, but the fields are 6-dimensional fields instead of
4-dimensional ones.
The gauge group is $G = $ SU(3) $\otimes$ SU(2) $\otimes  U(1)$ with corresponding fields $A_M = (G_M, W_M, B_M)$, where $G_M$ are
known as gluons, the gauge fields of $SU(3)$ gauge group, and $W_M, B_M$ are electroweak gauge bosons from $SU(2) \otimes U(1) $ gauge group.

We arrange the fermionic particles of three generations into doublets (left-handed spinors) and singlets (right-handed spinors),
included the right-handed neutrinos, as follows \cite{bars2006}
\begin{align*}
&\text{ The first generation} &
\begin{pmatrix}
u^L\\
d^L
\end{pmatrix}_{\fr{1}{3}}, \left(u^R\right)_{\fr{4}{3}}, \left(d^R\right)_{-\fr{2}{3}},
\begin{pmatrix}
\nu^L_e\\
e^L
\end{pmatrix}_{-1},\left(\nu^R_e\right)_0, \left(e^R\right)_{-2} ,
\\
&\text{The second generation} &
\begin{pmatrix}
c^L\\
s^L
\end{pmatrix}_{\fr{1}{3}}, \left(c^R\right)_{\fr{4}{3}}, \left(s^R\right)_{-\fr{2}{3}},
\begin{pmatrix}
\nu^L_\mu\\
\mu^L
\end{pmatrix}_{-1},\left(\nu^R_\mu\right)_0, \left(\mu^R\right)_{-2} ,
\\
&\text{The third generation} &
\begin{pmatrix}
t^L\\
b^L
\end{pmatrix}_{\fr{1}{3}}, \left(t^R\right)_{\fr{4}{3}}, \left(b^R\right)_{-\fr{2}{3}},
\begin{pmatrix}
\nu^L_\tau\\
\tau^L
\end{pmatrix}_{-1},\left(\nu^R_\tau\right)_0, \left(\tau^R\right)_{-2} .
\end{align*}
For convenience, we introduce the following notations:
\begin{gather}
(Q^{L_i})_\fr{1}{3}, (L^{L_i})_{-1} \text{ are respectively three left-handed quark and lepton doublets},\\
u^{R_i} = (u^R,c^R,t^R)_\fr{4}{3},
d^{R_i} = (d^R,s^R,b^R)_{-\fr{2}{3}},\\
e^{R_i} = (e^R,\mu^R,\tau^R)_{-2},
\nu^{R_i} = (\nu_e^R,\nu_\mu^R,\nu_\tau^R)_{0}\, ,
\end{gather}
 where $i = 1,2,3$.

The Lagrangian of the SM in 4+2 dimensions is given by
\[
L(A, \Psi^{L,R}, H, \Phi) = L(A) + L(A, \Psi^{L,R}) + L(\Psi^{L,R}, H) + L(A,H,\Phi)\, ,
\]
where  $L(A)$ is the Lagrangian for gauge bosons
\[
L(A) = -\dfrac{1}{4}Tr(G_{MN}G^{MN}) - \dfrac{1}{4}Tr(W_{MN}W^{MN}) -\dfrac{1}{4}B_{MN}B^{MN}\, .
\]
Here, the field strengths are defined as $A_{MN} = \partial_MA_N - \partial_NA_M - ig_A[A_M,A_N]$, $A= G, W, B$.

$L(A,\Psi^{L,R})$ describes the interaction between fermions and gauge boson via covariant derivatives
\begin{align}
L(A,\Psi^{L,R}) &= \dfrac{i}{2}\left(\bar{Q}^{L_i}\slashed{X}\slashed{\bar{D}}Q^{L_i}
  +
  \bar{L}^{L_i}\slashed{X}\slashed{\bar{D}}L^{L_i}
  \right)\\
&-\dfrac{i}{2}\left(\bar{d}^{R_i}\slashed{\bar{X}}\slashed{D}d^{R_i} +
 \bar{e}^{R_i}\slashed{\bar{X}}\slashed{D}e^{R_i}
 \right)\\
&-\dfrac{i}{2}\left(\bar{u}^{R_i}\slashed{\bar{X}}\slashed{D}u^{R_i} +
\bar{\nu}^{R_i}\slashed{\bar{X}}\slashed{D}\nu^{R_i}
\right) + h.c.
\end{align}

The Yukawa coupling
 $L(\Psi^{L,R}, H)$  has the  following structure
\bea
L(\Psi^{L,R}, H) &=& (g_u)_{ij}\bar{Q}^{L_i}\slashed{X}u^{R_j} H^c
+ (g_d)_{ij}\bar{Q}^{L_i}\slashed{X}d^{R_j} H\crn
&+& (g_\nu)_{ij}\bar{L}^{L_i}\slashed{X}\nu^{R_j} H^c
+ (g_e)_{ij}\bar{L}^{L_i}\slashed{X}e^{R_j} H + h.c.,\nn
\eea
where $H^c = i\si _2 H^*$ is the SU(2) charge conjugate of H. The Yukawa couplings $g_u, g_d, g_\nu, g_e$ are complex $3\times3$ constant matrices. It is possible to choose a basis for the quarks and leptons such that $g_u$ and $g_e$ are real and diagonal, while $g_d$ and $g_\nu$ are Hermitian but non-diagonal that relate to the Cabibbo - Kobayashi - Maskawa matrices for the quarks and neutrinos \cite{bars2006}.

The Higgs-Dilaton Lagrangian $L(A,\Phi,H)$ has the form:
\[
L(A,\Phi,H) = \dfrac{1}{2}\Phi\partial^2\Phi + \dfrac{1}{2}\Big(H^\dagger D^2H + (D^2H)^\dagger H \Big) - V(\Phi,H)\, ,
\]
where the covariant derivative is given by
\[
D_MH = \left(\partial_M - ig_2 W^a_M\dfrac{\si ^a}{2} - i\dfrac{g_1}{2}B_M\right)H\, .
\]

The Higgs-Dilaton  potential has the following form which is considered a result of b-gauge symmetry, the inspiration of this symmetry
comes from BRST formalism \cite{kuo2006} but it ultimately comes from the underlying $SP(2,R)$,
\[
V(\Phi,H) = \la\left(H^\dagger H - \al ^2 \Phi^2\right)^2 + V(\Phi)\, ,
\]
where $\la, \al $ are dimensionless couplings. $H$ and $\Phi$ are the SU(2) Higgs and Dilaton doublet respectively. The form of $V(\Phi)$ is not clear but it certainly plays a role in the construction of the effective potential in the next section.

\section{EWPT with Dilaton  in 2T Model}\label{sec3}

The Dilaton field in 2T physics plays an important role, being the source of the completed cosmic properties, including the electroweak phase transition \cite{1bars}.

To investigate the electroweak phase transition (EWPT) in the 2T model, we shall find out the effective potential constructed from the Lagrangian of the 2T model. However, the result should be shown in 1T-physics, i.e., 3+1 dimensions, which is familiar for us. This is the way we worked out:  First, we derive the 1T Lagrangian reduced from the 2T model by gauge fixing technique, this will be briefly introduced in the following section. Then we can construct the effective potential from the reduced 1T Lagrangian to examine the EWPT. The Dilaton field coming from the 2T model will contribute to this effective potential and will ensure a strong first-order phase transition satisfying the third Sakharov's condition about the baryon asymmetry of our Universe. Although the Dilaton mass is unknown, we could use the condition for the strength of phase transition to constraint the above-mentioned mass.

\subsection{Gauge fixing technique}

The gauge fixing technique was re-introduced, this technique plays an important role in reducing 2T model to 1T model. Because the 2T model with the $Sp(2, R)$ symmetry required the Higgs potential with an extra Dilaton field. If the Dilaton field thus takes a suitable form, the Dilaton field can become the trigger for the electroweak phase transition. To see this clearly, the gauge fixing technique is like a hologram that acts as a local symmetry. From there, reducing the Higgs-Dilaton 2T potential to 1T to investigate the activation of Dilaton.

By using the 2T gauge symmetry which is the two $SP(2,R)$ generators, in the worldline formalism $X^2 = (X.P + P.X) = 0$, we eliminate two components of $X^M$ and can obtain the 3+1 dimensional holographic image that emerges from the 4+2 dimensional system.

The flat metric in 4+2 dimensions is chosen by:
\begin{align}
&\eta_{MN} = \text{diag}(-1,-1,1,1,1,1),\qquad \text{where } M,N = {0',0,1',1,2,3}\, ,\\
\text{so } & X^2 = -(X^{0'})^2 - (X^0)^2 + (X^{1'})^2 + (X^1)^2 + (X^2)^2 + (X^3)^2 \,.
\end{align}
And the flat metric in $3+1$ dimensions is given by
\[
g_{\mu\nu} = \text{diag}(-1,1,1,1),
\qquad \text{where } \mu = 0,1,2,3\,.
\]

It is convenient if one chooses a lightcone basis in 4+2 dimensions written as
\begin{align}
&X^{\pm '}_i = \dfrac{1}{\sqrt{2}}\left(X^{0'}_i\pm X^{1'}_i\right), \text{where } X_i^M = (X^M_1,X^M_2) \equiv (X^M, P^M)\\
\Rightarrow&
\begin{cases}
X^2 = - 2X^{+'}X^{-'} + X^\mu X_\mu ,\\
X^MP_M = -X^{+'}P^{-'} - X^{-'}P^{+'} + X^\mu P_\mu\, .
\end{cases}
\end{align}

We choose the Bars's parametrization
 \cite{bars2000a} for the components of $X^M$ as follows
\begin{align}
X^{+'} = \ka , X^{-'} = \ka \La, X^\mu = \ka  x^\mu \label{3.1}\\
\Rightarrow \ka  = X^{+'}, \La = \dfrac{X^{-'}}{X^{+'}}, x^\mu = \dfrac{X^\mu }{X^{+'}}\, .
\end{align}
It is emphasized that this is one of the many possible choices to parameterize the $X_M$.

Using the above parametrization, one can get some useful operators in 4+2 dimensions and then apply them to reduce the kinematical equation to get the reduction of the field from 4+2 dimensions to 3+1 dimensions. The detailed derivation is shown in \cite{bars2006} so we just summarize the results below.
\begin{enumerate}[\textbullet]
\item For scalar fields, ones have got
\begin{align}\label{3.2a}
\begin{cases}
\Phi(X) \longrightarrow \dfrac{1}{\ka }\phi(x)\quad ;\quad H(X) \longrightarrow \dfrac{1}{\ka }h(x) , \\
\partial^2 \Phi(X) = \partial^M\partial_M \Phi(X) \longrightarrow \dfrac{1}{\ka ^3}\dfrac{\partial^2 \phi(x)}{\partial x^\mu \partial x_\mu}\, ,\\
D^2 H(X) = D^M D_M H(X) \longrightarrow \dfrac{1}{\ka ^3}D^\mu D_\mu h(x)\, .
\end{cases}
\end{align}
\item For chiral fermion fields, these reductions from $4+2$ dimensions to $3+1$ dimensions have been derived as
\begin{align}\label{3.2b}
\begin{cases}
\Psi^{L,R}(X) \longrightarrow \dfrac{1}{2^{1/4}\ka ^2}\begin{pmatrix}
\psi^{L,R}(x)\\
0
\end{pmatrix}\, ,\\
\bar{\Psi}^{L,R}(X)  \longrightarrow \dfrac{-i}{2^{1/4}\ka ^2}\begin{pmatrix}
0& \bar{\psi}^{L,R}(x)
\end{pmatrix}\, ,\\
g\bar{\Psi}^L\slashed{X}\Psi^R H \longrightarrow \dfrac{g}{\ka ^4}\bar{\psi}^L\psi^R h ,\\
g H^\dagger\bar{\Psi}^R\bar{\slashed{X}}\Psi^R \longrightarrow \dfrac{g}{\ka ^4}h^\dagger\bar{\psi}^R\psi^L
\end{cases}
\end{align}
\item The reductions of gauge boson field are written as
\begin{align}\label{3.2c}
\begin{cases}
A^{+'}(X) = A_{-'}(X) = 0 , \\
A^{-'}(X) =-A_{+'}(X) \longrightarrow \dfrac{1}{\ka }x^\mu A_\mu(x) ,\\
A^\mu(X) \longrightarrow \dfrac{1}{\ka }A_\mu(x)\, ,\\
F_{MN}(X)F^{MN}(X) \longrightarrow \dfrac{1}{\ka ^4}F_{\mu\nu}F^{\mu\nu}\,.
\end{cases}
\end{align}
\end{enumerate}

\subsection{Lagrangian reduction}

Since the effective potential is constructed in terms of Higgs background field, $\phi_c$, as well as Dilaton, which will be figured out to be identical with $\phi_c$ later, one should only consider the Lagrangian terms containing the Higgs field and Dilaton, then reduce it from 2T to 1T. So we are interested in the Higgs-Dilaton and Yukawa coupling terms in the Lagrangian.

Let consider the Yukawa terms first. We will only keep the top quark terms because its contribution dominates in the effective potential,
 while the other fermions are not heavy enough to significantly alter the result. The top quark terms extracted from Yukawa Lagrangian are given by
\[
L(\Psi^{L,R}, H) = (g_u)_{33}\bar{Q}^{L_3} \slashed{X} u^{R_3} H^c + (g_u^\dagger)_{33}H^{c\dagger} \bar{u}^{R_3}  \slashed{\bar{X}} Q^{L_3}\,.
\]
Using the result in the previous subsection (see  Eq.\eqref{3.2b}),  one can reduce the above Lagrangian from 2T to 1T as follows
\begin{align}
L(\Psi^{L,R}, H) \longrightarrow &-\dfrac{(g_u)_{33}}{\ka ^4} \bar{Q}^{L_3} u^{R_3} h^c -
 \dfrac{(g_u^\dagger)_{33}}{\ka ^4}h_c^\dagger \bar{u}^{R_3}   Q^{L_3}\\
=\ &-\dfrac{(g_u)_{33}}{\sqrt{2}\ka ^4} \begin{pmatrix}
\bar{t}^L & \bar{d}^L
\end{pmatrix}t^R\begin{pmatrix}
v+\si (x)\\
0
\end{pmatrix} - \dfrac{(g_u)_{33}}{\sqrt{2}\ka ^4} \begin{pmatrix}
v+\si (x) &0
\end{pmatrix}\bar{t}^R\begin{pmatrix}
t^L\\
d^L
\end{pmatrix}\\
=\ & -\dfrac{(g_u)_{33}}{\sqrt{2}\ka ^4} \ \bar{t}t (v+\si (x)) = - \dfrac{h_t}{\ka ^4}\ \bar{t}t\chi\,,
\end{align}
where we have used
\begin{gather}
h(x) = \dfrac{1}{\sqrt{2}}\begin{pmatrix}
0\\
v+\si (x)
\end{pmatrix} = \chi \begin{pmatrix}
0\\
1
\end{pmatrix}\qquad\text{in unitary gauge},\\
h^c = i\si ^2 h^* = \dfrac{1}{\sqrt{2}}i\begin{pmatrix}
0 &-i\\
i & 0
\end{pmatrix} \begin{pmatrix}
0\\
v+\si (x)
\end{pmatrix} = \dfrac{1}{\sqrt{2}}\begin{pmatrix}
v+\si (x)\\
0
\end{pmatrix},\\
g_u\text{ is Hermitian},\ g_u = g_u^\dagger,\ h_t \equiv (g_u)_{33}\,,\\
\bar{t}^Rt^L + \bar{t}^Lt^R = \bar{t}t.
\end{gather}

The more complicated work will be done with the Higgs-Dilaton  Lagrangian,
\[
L(A,H,\Phi) = \dfrac{1}{2}\Phi\partial^2\Phi + \dfrac{1}{2}\Big(H^\dagger D^2H + (D^2H)^\dagger H \Big) - \la\left(H^\dagger H -
\al ^2 \Phi^2\right)^2 - V(\Phi)\,,
\]
with the result in  Eq.\eqref{3.2a} from gauge fixing technique, one would be able to derive the following reduction
\begin{align}
L(A,H,\Phi) \longrightarrow \ &\dfrac{1}{2\ka ^4}\phi\dfrac{\partial^2\phi}{\partial x^\mu\partial x_\mu}
 + \dfrac{1}{2\ka ^4}\left[h^\dagger D_\mu  D^\mu h + (D^\mu D_\mu h)^\dagger h\right]\crn
& \qquad\qquad\qquad\qquad\qquad\qquad- \dfrac{\la}{\ka ^4}\left(h^\dagger h - \al ^2\phi^2\right)^2 - V(\phi)\,.
\end{align}
The first term is just  the kinetic term of the Dilaton. Our work is related to the later terms. For convenience, we introduce the following notation in the covariant derivative
\begin{align}
D_\mu h &= \left(\partial_\mu - ig_1W_\mu^a\dfrac{\si ^2}{2} - i\dfrac{g_2}{2}B_\mu\right) h\\
 &= \partial_\mu h - \dfrac{i}{2}\begin{pmatrix}
g_1W_\mu^3+g_2B_\mu & g_1W_\mu^+\sqrt{2}\\
g_1W_\mu^-\sqrt{2} & - g_1W_\mu^3+g_2B_\mu
\end{pmatrix} \\
&= \partial_\mu h - i \bm{P}_\mu\,,
\end{align}
where we have used $W^\pm_\mu = \fr{1}{\sqrt{2}} (W^1_\mu \mp iW^2_\mu)$.
We see that $\bm{P}_\mu$ is, as usual, Hermitian, i.e., $\bm{P}_\mu^\dagger = \bm{P}_\mu$. The gauge fields have the transformation rule:
\[
g_A A'_\mu(x) = g_ASA_\mu S^\dagger + iS(\partial_\mu S^\dagger)\,,
\]
and $S(x)$ is a matrix representation of gauge group of $A_\mu$, so $\bm{P}_\mu$ is equivalent to $\bm{P}'_\mu$ as follows
\[
\bm{P}'_\mu = \xi \bm{P}_\mu \xi^\dagger + i\xi(\partial\xi^\dagger)\,,
\]
with $\xi$ is any unitary matrix. Expanding the covariant derivative yields
\begin{align}
h^\dagger D_\mu  D^\mu h + (D^\mu D_\mu h)^\dagger h &= h^\dagger(\partial_\mu - i\bm{P}_\mu)(\partial^\mu - i\bm{P}^\mu) h +
\left[(\partial_\mu - i\bm{P}_\mu)(\partial^\mu - i\bm{P}^\mu)h \right]^\dagger h\\
&= h^\dagger (\partial^2 h) + (\partial^2 h^\dagger)h\\
 &- 2ih^\dagger \bm{P}^\mu \partial_\mu h + 2i(\partial_\mu h^\dagger)\bm{P}^\mu h\\
 &-2h^\dagger\bm{P}^2 h\,.
\end{align}
Ignoring unitary gauge for a moment,  one can rewrite  the Higgs field  as
\[
h = \chi\bm{\zeta}\bm{\va}_0 = \chi\begin{pmatrix}
\zeta_1^* & \zeta_2\\
\zeta_2^* &\zeta_1
\end{pmatrix}\begin{pmatrix}
0\\
1
\end{pmatrix}\,,
\]
where $|\zeta_1|^2 + |\zeta_2|^2 = 1$ so $\bm{\zeta}$ is an unitary matrix and $h^\dagger h = \chi^2$. Substitute this Higgs field, one gets
\begin{align}
h^\dagger (\partial^2 h) + (\partial^2 h^\dagger)h &= 2\chi\partial^2\chi - 2\chi^2\bm{\va}_0^\dagger
(\partial_\mu \zeta^\dagger)(\partial^\mu \zeta)\bm{\va}_0,\\
- 2ih^\dagger \bm{P}^\mu \partial_\mu h + 2i(\partial_\mu h^\dagger)\bm{P}^\mu h & = -2i\chi^2\bm{\va}_0^\dagger\zeta^\dagger
 \bm{P}^\mu (\partial_\mu \zeta)\bm{\va}_0 + 2i\chi^2\bm{\va}_0^\dagger(\partial_\mu \zeta^\dagger) \bm{P}^\mu \zeta\bm{\va}_0,\\
-2h^\dagger\bm{P}^2 h & = -2\chi^2\bm{\va}_0^\dagger\zeta^\dagger\bm{P}^2\zeta\bm{\va}_0\,.
\end{align}
After collecting those results and simplifying, one obtains
\begin{align}
h^\dagger D_\mu  D^\mu h + (D^\mu D_\mu h)^\dagger h &= 2\chi\partial^2\chi - 2i\chi^2 \bm{\va}_0^\dagger
( i \zeta^\dagger \partial_\mu \zeta + \zeta^\dagger \bm{P}_\mu\zeta)\zeta^\dagger\partial^\mu\zeta \bm{\va}_0 \\
&- 2\chi^2 \bm{\va}_0^\dagger(i\zeta^\dagger\partial_\mu\zeta + \zeta^\dagger \bm{P}_\mu\zeta)\zeta^\dagger
\bm{P}^\mu\zeta\bm{\va}_0\\
&= 2\chi\partial^2\chi -2\chi^2\bm{\va}_0^\dagger\bm{P}_\mu (i\zeta^\dagger\partial^\mu\zeta + \zeta^\dagger\bm{P}^\mu\zeta)
\bm{\va}_0\\
&= 2\chi\partial^2\chi -2\chi^2\bm{\va}_0^\dagger\bm{P}^2\bm{\va}\\
&= 2\chi\partial^2\chi - \dfrac{\chi^2}{2}(g_1^2 + g_2^2)Z_\mu Z^\mu - g_1^2\chi^2 W^-_\mu W^{\mu,+}\,.
\end{align}
To derive the final line, we have used
\begin{align}
\bm{\va}_0^\dagger\bm{P}^2\bm{\va} &= \dfrac{1}{4}\begin{pmatrix}
0 &1
\end{pmatrix}\begin{pmatrix}
g_1W_\mu^3+g_2B_\mu & g_1W_\mu^+\sqrt{2}\\
g_1W_\mu^-\sqrt{2} & - g_1W_\mu^3+g_2B_\mu
\end{pmatrix}\crn
&\qquad\qquad\qquad\qquad\times\begin{pmatrix}
g_1W_\mu^3+g_2B_\mu & g_1W_\mu^+\sqrt{2}\\
g_1W_\mu^-\sqrt{2} & - g_1W_\mu^3+g_2B_\mu
\end{pmatrix}\begin{pmatrix}
0\\
1
\end{pmatrix}\\
&= \dfrac{1}{2}g_1^2 W^-_\mu W^{\mu,+} + \dfrac{1}{4}(- g_1W_\mu^3+g_2B_\mu)^2\\
&= \dfrac{1}{2}g_1^2 W^-_\mu W^{\mu,+} + \dfrac{g_1^2+g_2^2}{4}Z_\mu Z^\mu\,,
\end{align}
where we have defined new fields $Z_\mu, A_\mu$ that transform from the older fields $W_\mu^3, B_\mu$
\[
\begin{cases}
Z_\mu = \cos\theta_W W^3_\mu - \sin\theta_W B_\mu\\
A_\mu = \sin\theta_W W^3_\mu + \cos\theta_W B_\mu\, .
\end{cases}
\]
The Weinberg angle or weak angle is defined as
\begin{gather}
\tan\theta_W =
  \dfrac{g_1}{g_2}\,.
\end{gather}

Finally, one can rewrite the reduction of Higgs-Dilaton  Lagrangian as follows
\begin{align}
L(A,H,\Phi) &\longrightarrow \chi\partial^2\chi - \dfrac{g_1^2 + g_2^2}{4}\chi^2Z_\mu Z^\mu - \dfrac{g_2^2}{2} \chi^2 W^-_\mu W^{\mu,+}\crn
&\qquad\qquad\qquad\qquad\qquad\qquad\qquad- \la\left(\chi^2 - \al ^2\phi^2\right)^2 - V(\phi)\,.
\end{align}

From the Lagrangian reduction, one can get the mass term for the fields in terms of Higgs background field $\phi_c$. Let us  redefine
the Higgs component $\chi$ as
\[
\chi(x) = \dfrac{1}{\sqrt{2}}(\phi_c + h(x))\,,
\]
where $\phi_c$ is the Higgs background field and $h(x)$ is the Higgs field. Then one can deduce the masses of top quark and gauge bosons at tree level as follows
\begin{align}
m_t^2(\phi_c) &= \dfrac{h_t^2}{2}\phi_c^2 = \dfrac{m_t^2}{v^2}\phi_c^2\,,\\
m_W^2(\phi_c) &= \dfrac{g^2_1}{4}\phi_c^2  = \dfrac{m_W^2}{v^2}\phi_c^2\,,\\
m_Z^2(\phi_c) &= \dfrac{g^2_1+g_2^2}{4}\phi_c^2 = \dfrac{m_Z^2}{v^2}\phi_c^2\,,
\end{align}
where $m_i \equiv m_i(v)$ are the mass of field $i$ at vacuum and we have absorbed the coefficient $\ka$ into vacuum masses.

For the masses of Higgs and Dilaton, one could expand the Higgs-Dilaton  potential
\begin{align}\label{47v}
V(\Phi,H) &= \la\left(H^\dagger H - \al ^2 \Phi^2\right)^2 + V(\Phi)\\
\longrightarrow\  V(h,\phi) &= \la\left[\chi^2 - \al ^2\phi^2\right]^2 + V(\phi)\,.
\end{align}

The 2T-gauge symmetry requires the above Higgs-Dilaton potential to be purely quartic, so it adds to the Dilaton component. The intensity of interaction between Dilaton and Higgs is $\alpha$. Therefore the VEV of Dilaton is directly proportional to the VEV of the Higgs field by $\alpha$ (for details, see  Ref. \cite{bars2006}).  This potential is remarkable because the quadratic mass term of  the Higgs field does not appear, or in other words, is forbidden. This is also related to the b-symmetry as we have mentioned. One can understand the reason for the pure quartic interaction directly by examining the equation of motion for the interacting scalar fields \cite{bars2006}.

Therefore, in $4+2$ dimensions or 2T-physics, there is no mechanism like the dynamical breakdown of the $SU(2)\otimes U(1)$ electroweak symmetry with a tachyonic mass term for the Higgs field $H$ to generate the mass for particle via VEV of Higgs boson. However, the coupling to the Dilaton generates the non-trivial vacuum configuration $H^\dagger H - \al ^2 \Phi^2 = 0$ with $V(\Phi)=0$ \cite{bars2006}.
 Hence the equations of motion for scalar are
\[
\partial^2 H(X) = 0 \quad,\quad \partial^2\Phi = V'(\Phi)\,.
\]
 We have $\partial^2 H(X) = \ka ^{-3}\partial_x^2 h(x)$ as in Eq.\eqref{3.2a}, so $h(x)$ must be a constant at the vacuum, i.e., the VEV $v$, but still depends on $\ka $ as follows
\[
\langle H(X) \rangle = \dfrac{v}{\ka }\begin{pmatrix}
0\\
1
\end{pmatrix}\,.
\]
On the other hand, the Dilaton  field $\Phi(X)$ must not only satisfy the vacuum configuration $H^\dagger H - \al ^2 \Phi^2 = 0$, but also
be homogeneous as considering in the preceding part, so
\[
\langle\Phi(X)\rangle = \pm \dfrac{v}{\al \ka }\,.
\]
We can think the way of taking the value $\langle\Phi\rangle \neq 0$ is to fit the phenomenology of the Higgs $\langle H\rangle \neq 0$ and $\langle\Phi\rangle$ that might be stabilized by additional interactions in the gravitational or string theory to have a fixed value related to VEV of Higgs $v$, extra dimension $\ka $ and the coupling $\al $. Therefore, the Dilaton that appeared in the  Higgs-Dilaton potential plays the same role as the tachyonic mass term of Higgs boson in usual SM and can be considered as the source leading to the electroweak symmetry breaking \cite{bars2006}.

As a result, since the VEVs of Higgs and Dilaton are not really different, only for the coefficient of coupling constant $\al$ contributed to the Dilaton, we could infer that their background field should be considered simultaneously and one is enough, i.e., Higgs background field. This is completely a consequence of 2T model and its symmetry. On the other hand, there are some models that also considered the contribution of a real singlet $S(x)$ to effective potential (for example, see Refs \cite{egkr2012,ck2013}). In the mentioned models, the Higgs and Dilaton background fields are considered to be distinct because the Higgs-Dilaton potential would not be constrained uniquely like in the 2T model.

However, by taking the potential in Eq.(\ref{47v}) we obtain the breaking of the electroweak symmetry through the VEV of the Dilaton $\Phi$.  Thus the breaking of the electroweak symmetry and the obtained VEV of Dilaton must go together \cite{bars2001}. If $\al = 0$, then there is no interaction between Higgs and Dilaton, and in this case, the Dilaton gains  VEV first then Higgs breaks the electroweak symmetry later. The Dilaton-driven electroweak phase transition makes a lot more sense conceptually than the usual way \cite{bars2001}.

Now we choose again the unitary gauge for the Higgs and absorb three of its degrees of freedom into the electroweak gauge fields, i.e., to the Z and W bosons. We expand Higgs and Dilaton around their background field, which  is really identical as we have discussed above. The remaining neutral Higgs and the Dilaton  field reductions can be written as follows
\[
H^0(X) \longrightarrow \dfrac{1}{\ka }\chi(x) = \dfrac{1}{\ka  \sqrt{2}}(\phi_c + h(x))\quad,\quad \Phi(X) \longrightarrow
\dfrac{1}{\al \ka \sqrt{2}}(\phi_c + \al  d(x)),
\]
here $d(x)$ is the Dilaton field. As we have mentioned when we derive the field equation for interacting scalar field in the above section, the quadratic mass terms are forbidden in the Higgs-Dilaton potential in 4 + 2 dimensions (d = 4) while only the quartic terms are allowed \cite{bars2006}. This is related to the b-symmetry that ultimately comes from the underlying Sp(2,R)\cite{bars2006}.  Let us  assume the extra potential has the form
\begin{equation}
V(\Phi) = \rho \Phi^4 - \dfrac{\om ^2}{\ka ^2}\Phi^2 \longrightarrow V(\phi) = \dfrac{\rho}{\ka ^4}\phi^4 -
\dfrac{\om ^2}{\ka ^4}\phi^2,
\end{equation}
where $\rho, \om $ are coupling constants. $\dfrac{\om ^2}{\ka ^2}\Phi^2$ as an effective term which comes from another interaction of Dilaton (For example, the interaction between the Dilaton and the additional $S(x)$ field is suggested in Refs.\cite{1bars,2bars,2barsb}). We accept that this term must be very small (ie $\omega^2/\kappa^2\Phi^2$ is very less than $\rho\Phi^4$) or a softly broken $SP(2, R)$ symmetry. The most important role is to minimize the Higgs-Dilaton potential, without this component the Higgs-Dilaton potential cannot be minimized with only the component $\rho \phi^4$.

We have minimized again the potential Eq.~\ref{47v}. In the condition $V(\Phi)=0$, the potential Eq.~\ref{47v} will have a minimum condition like $H^\dagger H - \alpha^2 \Phi^2 =0$. Therefore, using this condition ($H^\dagger H - \alpha^2 \Phi^2 =0$) as a limiting condition for the minimization of Eq.~\ref{47v}.

This assumption could be justified if one takes into account the next leading order of 2T metric in the action \cite{bars2006} and $Z_2$ symmetry of Dilaton. Then by substituting the above Higgs and Dilaton fields into the Higgs-Dilaton potential, we can compute
\begin{align} \label{50}
V(h,\phi) &= \dfrac{\lambda}{4}\left[(\phi_c + h)^2 - (\phi_c + \alpha d)^2\right]^2 - \dfrac{\omega^2} {2\alpha^2}(\phi_c + \alpha d)^2 + \dfrac{\rho}{4\alpha^4}(\phi_c + \alpha d)^4 \nonumber\\
&= \dfrac{\lambda}{4}(h-\alpha d)^2 (2\phi_c + h +\alpha d)^2- \dfrac{\omega^2} {2\alpha^2} (\phi_c^2 + 2\alpha\phi_c d + \alpha^2 d^2)\nonumber \nonumber\\
&\qquad\qquad\quad + \dfrac{\rho}{4\alpha^4}(\phi_c^4 + 4\alpha\phi_c^3 d + 6\alpha^2\phi_c^2 d^2 +4\alpha^3 \phi_c d^3 + \alpha^4d^4) \nonumber\\
&= \dfrac{\lambda}{4}(4\phi_c^2 h^2 + 4\alpha^2\phi_c^2 d^2- 8\alpha\phi_c^2 h d) - \dfrac{\omega^2} {2\alpha^2} (\phi_c^2 + 2\alpha\phi_c d + \alpha^2 d^2)\nonumber \nonumber\\
&\qquad\qquad\quad + \dfrac{\rho}{4\alpha^4}(\phi_c^4 + 4\alpha\phi_c^3 d + 6\alpha^2\phi_c^2 d^2) + \text{3,4-fields interaction terms} \nonumber\\
&=\left\{\left(\dfrac{\rho}{4\alpha^4}\phi_c^4 - \dfrac{\omega^2}{2\alpha^2}\phi_c^2\right) + \left(\dfrac{\rho}{\alpha^3}\phi_c^3-\dfrac{\omega^2 }{\alpha}\phi_c\right)d\right.\nonumber \nonumber\\
&\qquad\left.-2\lambda\alpha\phi^2_chd + \lambda\phi_c^2 h^2 +\left(\lambda\alpha^2\phi_c^2 - \dfrac{\omega^2}{2} + \dfrac{3\rho}{2\alpha^2}\phi_c^2 \right)d^2\right\} + \text{interaction terms},
\end{align}
from which we can deduce (in the following we will neglect the $\kappa$ dependence overall factor for a moment) the tree-level potential
\begin{equation}
V_0(\phi_c) = \dfrac{\rho}{4\alpha^4}\phi_c^4 - \dfrac{\omega^2}{2\alpha^2}\phi_c^2, \nonumber
\end{equation}
which extremum satisfy the condition
\begin{equation}
V'_0(\phi_c) = \dfrac{\rho}{\alpha^4}\phi_c^3-\dfrac{\omega^2 }{\alpha^2}\phi_c = 0, \nonumber
\end{equation}
This condition exactly eliminates the linear term with respect to $d(x)$ appearing in Eq.(\ref{50}). If there is no $\omega^2 \Phi^2/\kappa^2$, to minimize the Higgs-Dilaton potential, in Eq.\ref{50}, we force $\rho=0$. Solving the above condition, we have two solutions, one is trivially $\phi_c = 0$ and the other is $\phi_c^2 = \omega^2 \alpha^2/\rho$. Since $\alpha$ is a  real coupling constant, the later solution is only valid when $\omega^2/\rho > 0$, and one can choose $\rho > 0, \omega^2 > 0$ to make the extremum go to the minimum while the extremum at $\phi_c = 0$ becomes the maximum.  The solution of $\phi_c$ corresponding to the minimum will be called by VEV of Higgs field at $0K$, which we have denoted by $v$,
\[
v^2 = \dfrac{\omega^2\alpha^2}{\rho} \, .
\]

This potential should have a minimum at $h(x) = d(x) = 0$, in case of VEVs take place. Consequently, all linear terms of $h$ or $d$ have to disappear. To ensure the linear term of $d(x)$ vanishes in Eq.(\ref{50}), we  propose $\om ^2 \approx \fr{\rho}{\al ^2}\phi_c^2$, which also makes sure the vanishing of  the masses of Goldstone bosons. After diagonalization, we obtain physical particles which have masses at tree level as
\begin{align}
m^2_{d'}=&\frac{2 \alpha ^4 \lambda  \phi_c^2+2 \alpha ^2 \lambda  \phi_c^2+3 \rho  \phi_c^2-\alpha ^2 \omega ^2}{2 \alpha ^2}\nonumber\\
&-\frac{\sqrt{8 \alpha ^2 \lambda  \phi_c^2 \left(\alpha ^2 \omega ^2-3 \rho  \phi_c^2\right)+\left(\phi_c^2 \left(2 \alpha ^4 \lambda +2 \alpha ^2 \lambda +3 \rho \right)-\alpha ^2 \omega ^2\right)^2}}{2 \alpha ^2},\label{gt1}\\
m^2_{h'}=&\frac{2 \alpha ^4 \lambda  \phi_c^2+2 \alpha ^2 \lambda  \phi_c^2-\alpha ^2 \omega ^2+3 \rho  \phi_c^2}{2 \alpha ^2}\nonumber\\
&+\frac{\sqrt{8 \alpha ^2 \lambda  \phi_c^2 \left(\alpha ^2 \omega ^2-3 \rho  \phi_c^2\right)+\left(\phi_c^2 \left(2 \alpha ^4 \lambda +2 \alpha ^2 \lambda +3 \rho \right)-\alpha ^2 \omega ^2\right)^2}}{2 \alpha ^2}.\label{gt2}
\end{align}

In the equation \ref{gt1}, as $\omega$ and $\rho$ go to zero, we obtain $m^2_d=0$, which is in agreement with the results in the Ref. \cite{bars2006}. The above formulas can be rewritten as follows

\begin{align}
m_{h'}^2(\phi_c) &= A \phi_c^2 ,, \\
m_{d'}^2(\phi_c) &= B\phi_c^2\,,
\end{align}
here $A,B$ are the parameteres. For convenience, $h',d'$ will be denoted again as $h,d$ in the following sections. The contribution of extra dimensions does not only appear in one effective potential via Dilaton but also in the mass of Higgs and other fields as well.

Note that all the $\kappa$ coefficients will be simplified by the scaling invariant of the actions through $\delta(X^2)$ when reducing from 2T to 1T \cite{bars2006,kuo2006}. Because of this, in 2T, $\omega/\kappa<\rho$ or $\kappa$ is very large, or $\omega^2/\kappa^2\Phi^2\ll \rho\Phi^4$. However, when reducing from 2T to 1T, the parameter $\kappa$ is suppressed, so in 1T, $\omega^2\phi^2$, which is comparable to $\rho\phi^4$, plays the more important role. Thus, in 2T, the dilaton as a degree of freedom (since the component $\omega^2\phi^2$ is so small, it can be ignored leading to the dilaton mass in 2T being close to zero \cite{bars2006}). But in 1T, Dilaton has a mass that can be very different from zero.

\subsection{The effective potential for Higgs field and EWPT}

The symmetry-breaking process generating the mass of SM-like particles can take place when the VEV of the Higgs field jumps from zero to non-zero. We can see if this transition is due to the fact that the Dilaton field also jumps from zero to non-zero, or in other words, the VEV of the Higgs is anchored to the VEV of the Dilaton field. However, the mass generation mechanism for SM-like particles is due to the Higgs field, since there is only directly Yukawa interaction between the Higgs and SM-like particles in this model. Therefore, we consider a 1T effective Higgs potential as an effective component of the Higgs-Dilaton 2T potential.

The effective potential can be constructed by many methods, but the final result is the same. A well-known method is  a functional approach with the one-loop approximation, which is firstly introduced by Coleman and Weinberg and has been developed by Jackiw \cite{cjd1,cjd2,cjd3}. The effective potential could be constructed at zero temperature and then we have to add the thermal contribution to the final result. The one-loop effective potential at zero temperature can be written as
\[
V^{0K}_\text{eff}(\phi_c) = \dfrac{\la_R}{4}\phi_c^4 - \dfrac{m_R^2}{2}\phi_c^2 + \La_R + \dfrac{1}{64\pi^2}
\sum\limits_{i = h,d,W,Z,t} n_i m_i^4(\phi_c)\ln \dfrac{m^2_i(\phi_c)}{v^2}\,,
\]
where $\la_R, m_R, \Om_R$ are renormalized parameters,  $m_i(\phi_c)$ is mass at tree level derived in previous subsection, i.e., SM particles and Dilaton;
$n_i$ with $i = d,h,W,Z,t$ are constants related to the degrees of freedom of each field and are given by
\[
n_h = n_d = 1, n_W = 6, n_Z = 3, n_t = -12 \, .
\]

To work out renormalized parameters, we will use familiar normalization conditions,
\[
\begin{cases}
V_\text{eff}(v) = 0 \, ,\\
V'_\text{eff}(v) = 0\, ,\\
V''_\text{eff}(v) = m_h^2\,,
\end{cases}
\]
from which we find out the following results
\[
\begin{cases}
\la_R = \dfrac{m_h^2}{2v^2} - \dfrac{1}{16\pi^2 v^4}\sum\limits_{i} n_i m_i^4 \left(\ln\dfrac{m_i^2}{v^2}+\dfrac{3}{2}\right)\, , \\
m_R^2 = \dfrac{m_h^2}{2} - \dfrac{1}{16\pi^2v^2}\sum\limits_i n_i m_i^4\, , \\
\La_R = \dfrac{m_h^2v^2}{8} - \dfrac{1}{128\pi^2}\sum\limits_i n_i m_i^4\, .
\end{cases}
\]

After adding the thermal contribution, the effective potential becomes
\begin{align}
V_\text{eff}(\phi_c,T) &= \dfrac{\la_R}{4}\phi_c^4 - \dfrac{m_R^2}{2}\phi_c^2 + \La_R + \dfrac{1}{64\pi^2}
\sum\limits_{i = h,d,W,Z,t} n_i m_i^4(\phi_c)\ln \dfrac{m^2_i(\phi_c)}{v^2}\crn
& +\dfrac{T^4}{2\pi^2}\left\{\sum\limits_{i = h,d,W,Z} n_iJ_B[m_i^2(\phi_c)/T^2] + n_tJ_F[m_t^2(\phi_c)/T^2]\right\}\,,
\end{align}
where $J_B(m^2/T^2)$ and $J_F(m^2/T^2)$ are the thermal bosonic and fermionic functions, respectively, and are defined as follows
\[
J_k(m^2\beta^2) = \int\limits_0^\infty dx\ x^2 \ln\left[1+n_k e^{-\sqrt{x^2+\beta^2m^2(\phi_c)}}\right], \quad\text{ with }n_F = 1, n_B = -1 \, .
\]
At very high temperature, i.e., $\beta = T^{-1}$ is small, one can take the expansions given by \cite{quiros1999}

\begin{align}
J_B(m^2\beta^2) =& -\dfrac{\pi^4}{45} + \dfrac{\pi^2}{12}\dfrac{m^2}{T^2} - \dfrac{\pi}{6}\left(\dfrac{m^2}{T^2}\right)^{3/2} - \dfrac{1}{32}\dfrac{m^4}{T^4}\ln\dfrac{m^2}{a_bT^2}\\
&-2\pi^{7/2}\sum\limits_{n=1}^\infty(-1)^n \dfrac{\zeta(2n+1)}{(n+1)!},\\
J_F(m^2\beta^2) =& \dfrac{7\pi^4}{360} - \dfrac{\pi^2}{24}\dfrac{m^2}{T^2} - \dfrac{1}{32}\dfrac{m^4}{T^4}\ln\dfrac{m^2}{a_fT^2}\\
&-\dfrac{\pi^{7/2}}{4}\sum\limits_{n=1}^\infty(-1)^n \dfrac{\zeta(2n+1)}{(n+1)!}(1-2^{-2n-1})\,,
\end{align}
where $a_b = 16\pi^2\exp(3/2-2\ga _E)$, $a_f = \pi^2\exp(3/2-2\ga _E)$ which imply $\ln a_b
\approx 5.4076$, $\ln a_f \approx 2.6351$ and $\zeta$
 is the Riemann $\zeta$-function.

Expanding the thermal function and neglecting the field independent terms and $\zeta$-function terms, one can rewrite the effective potential as follows
\[
V_\text{eff}(\phi_c,T) = \dfrac{\la(T)}{4}\phi_c^4 - ET\phi_c^3 + D(T^2-T^2_0)\phi_c^2,
\]
\begin{gather}
\la(T) = \dfrac{m_h^2}{2v^2} + \dfrac{1}{16\pi^2v^4}\left(\sum\limits_{i = h,d,W,Z}n_i m_i^4\ln\dfrac{A_bT^2}{m_i^2}+ n_t m_t^4\ln\dfrac{A_fT^2}{m_t^2}\right),\\
E = \dfrac{m_h^3+m_d^3+6m_W^3 + 3m_Z^3}{12\pi v^3},\\
D = \dfrac{m_h^2+m_d^2+ 6m_W^2 + 3m_Z^2 +6m_t^2}{24v^2},\\
T_0^2 = -\dfrac{m_h^2}{4} + \dfrac{1}{32\pi^2v^2}\sum\limits_{i=h,d,W,Z,t} n_i m_i^4\,.
\end{gather}

The value of $V_{eff}$ at two minima become equal at the critical temperarure. Therefore, we only need to calculate a 2nd minimum and set the value of the effective potential at the 2nd minimum equal to zero because the value of the effective potential at the zero minimum is equal to zero, we can derive the critical temperature,
\[
T_c =  \dfrac{T_0}{\sqrt{1 - \dfrac{E^2}{\la(T_c)D}}}\,.
\]

This effective potential has only one unknown parameter - the Dilaton mass. We choose $m_d = 400 $ GeV, the others are given in Tab. \ref{tab:3.1}.

\begin{table}[htbp]
\begin{center}
\begin{tabular}{m{2cm}|m{2cm}|m{2cm}|m{2cm}|m{2cm}|m{2cm}}
\hline\hline
$m_h$ &$m_d$& $m_W$ & $m_Z$ & $m_t$ & $v$\\
\hline
125.09 &400& 80.385 & 91.1876 & 173.1 & 246\\
\hline\hline
\end{tabular}
\caption{The masses of particles and the VEV of the Higgs field in units of GeV.}
\label{tab:3.1}
\end{center}
\end{table}

\begin{figure}[htbp]
\centering
\includegraphics[width = 0.5\textwidth]{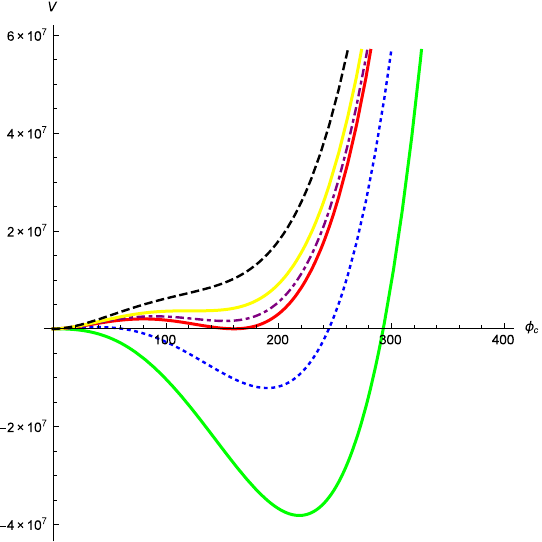}
\caption{The effective potential in the SM
deduced from 2T-physics to 1T-physics with different temperature. The black dash line: $T=130$ GeV. The yellow line: $T=T_1=125.811$ GeV. The purple dashed line: $ T=124$ GeV. The red line:  $T=T_c=122.79$ GeV. The blue dashed line:  $T=115$ GeV and the green line: $T=T_0=103.632$ GeV.}
\label{fig:3.1}
\end{figure}

The effective potential drawn in Fig. \ref{fig:3.1} shows a strong first-order phase transition with a high barrier potential between two local minima at appropriate range of temperature, for this case from $T_1 = 125.811$ GeV to $T_c = 122.79$ GeV which agrees with the energy range of EWPT that we expected ($v \sim 100$ GeV). $T_1$ is the temperature at which the effective potential begins to have a non-zero second minimum.

This result is different from the EWPT of the SM in the usual 1T-physics, which only implies a weak first-order phase transition. The contribution of the Dilaton to the effective potential, particularly the factor of cubic term $E$, acts as a trigger for a strong first-order phase transition. In other words, the extra dimensions from the 2T model would alter the result of the physics in 1T-physics via the Dilaton. The other results deduced from Fig. \ref{fig:3.1} is the same as that we have learned from 1T-physics. At  a very high temperature, the second minimum disappears and there is only a minimum at the origin which implies the symmetry restoration and the gauge bosons and fermions are massless. When the temperature drops below the critical temperature $T_c$, the system switches to the symmetry-breaking phase and the VEV takes a place at the vicinity of 220 GeV, which is not in agreement with the measurement $v = 246$ GeV, but this result could be acceptable because the expansions of thermal function are only reliable at high temperature.

\begin{figure}[h]
\centering
\includegraphics[width =0.8\textwidth]{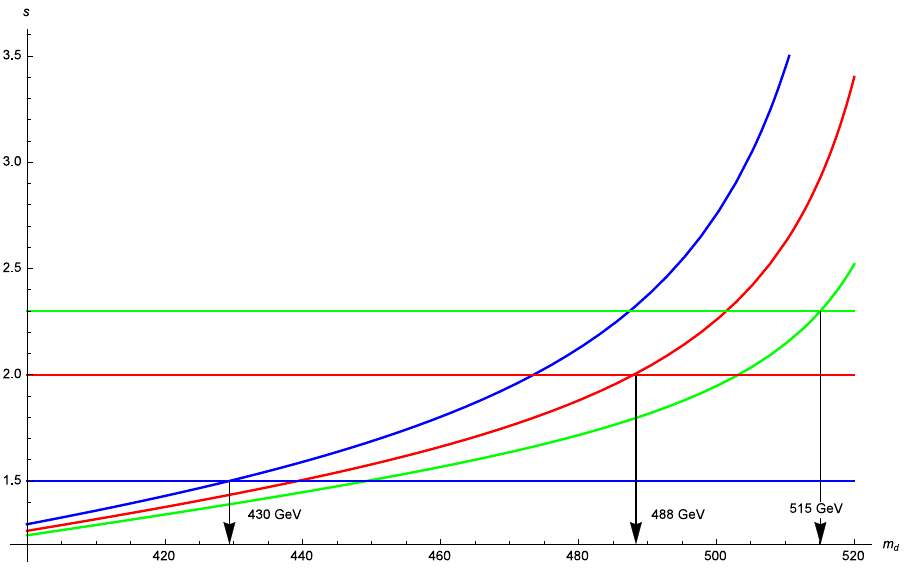}
\caption{The solution for the Dilaton mass with the strength of phase transition $s = 1.5,2,2.3$, the colors represented for
 each values of $s$ are blue, red, green, respectively.}
\label{fig:3.2}
\end{figure}
Next, we will work out the bound for the mass of Dilaton. The strength of the phase transition is given by \cite{phonglongvan1,phonglongvan2,phonglongvan3,phonglongvan4}
\[
s = \dfrac{2E}{\la(T_c)},
\]
where $\la(T_c)$ is calculated by substituting the critical temperature
\[
T_c =  \dfrac{T_0}{\sqrt{1 - \dfrac{E^2}{\la(T_c)D}}} = \dfrac{T_0}{\sqrt{1 - s\dfrac{E}{2D}}}\,.
\]

This may be a problem because $\la(T_c)$ depends on $s$, so the equation for $s$ is self-consistent. To evaluate  result numerically, we will use a graphic method: by choosing a value of $s$ for the expression of $T_c$, we draw a function of $s = 2E/\la(T_c)$
respecting to the mass of the Dilaton and point out the intersection of two figures to get the value of the Dilaton mass.  In order to have first-order phase transition, the strength $s$ must be larger than  unity, i.e., $s \geq 1$. Therefore, we will start with $s = 1$ and then increase it to have more results. Some results obtained are shown in Fig. \ref{fig:3.2}. By using this method, we found the strength of the EWPT in 2T model is in the range $1\leq s < 3.08$ and the Dilaton  mass   is approximately in the range $[345 \div 625]$ GeV.
  The result is summarized in Tab. \ref{tab:3.2}.

\begin{table}[h]
\begin{center}
\begin{tabular}{m{2cm}|m{2cm}|m{2cm}|m{2cm}|m{2cm}|m{2cm}}
\hline\hline
$s=\frac{2E}{\lambda_{T_C}}$&$1$&1.5&2&2.3&$3.08$\\
\hline
$m_d [GeV]$ &344.4& 430 & 488 & 515 & 625\\
\hline\hline
\end{tabular}
\caption{The results of Dilaton mass with different values of the strength of phase transition.}
\label{tab:3.2}
\end{center}
\end{table}

\section{Conclusion and discussion}\label{sec5}

The symmetry-breaking process in the 2T model happens with two steps. The first step is breaking the $SP(2,R)$ symmetry, reducing the 2T spacetime to 1T spacetime. Then there is the symmetry breaking like the SM one.
\begin{align*}
\text{2T model}&\\
\Downarrow& \text{breaking SP(2,R)}\\
\text{1T SM}&\\
\Downarrow& \text{breaking SU(2)} \\
\text{QED: }U(1)_Q&
\end{align*}

The structure of EWPT in the 2T model with the effective potential at finite temperature has been drawn at the 1-loop level, this potential only has one stage. The first-order EWPT is trigged by Dilaton.

The addition of an extra scalar particle that can trigger a first order electroweak phase transition, is not a new problem. The key issue, however, is the mechanism for adding the extra scalar particle. In the 2T model, the expansion of the space-time due to the symmetry of $Sp(2,R)$ led to the natural appearance of Dilaton. Then, according to the analysis in this paper, the dilaton will trigger a first order electroweak phase transition. Therefore, we can see that if the space-time is viewed as 6 dimensions, it will lead to an electroweak phase transition as expected.

These results can be said that the extra dimension has indirect contributions to the EWPT. One intriguing feature of the 2T extension is that it can provide a suitable picture of the matter-antimatter asymmetry.

The new source for EWPT is the Dilaton interacting with Higgs, so the Higgs-Dilaton potential in this model may be similar to that of composite Higgs. Future experimental studies of Higgs can reveal more composite particles and the Dilaton is a candidate.

Neutrino mass in the 2T model can be generated by a 10-dimension operator which may be a source of CP violation. Therefore, the effective potential survey in higher-order approximation will make the EWPT strength higher. This is an interesting work in the near future in a 2 or 3-loops effective potential.

Although Dilaton can explain EWPT, we still have very little information about it. For more predictions for Dilaton, ee can combine the EWPT with the cosmic inflation. If the Dilaton potential has a quadratic term that can be used by the standard inflation, we must introduce a dimensionful mass parameter. But that will break the basic $SP(2,R)$ symmetry. However, the parameter can be retrieved from the coupling between Dilaton and an additional scalar $S(x)$ which plays a role like the Dilaton, as discussed in Refs.\cite{1bars,2bars,2barsb}. This way was done consistently with the required local scale invariance demanded by 2T-physics Ref.\cite{3bars}. This is a follow-up research direction after this article.

Going beyond matter-antimatter symmetry. The 2T model can be reduced to the Randall-Sundrum model (5 dimensions) but we must not accept the $SP(2,R)$ symmetry. The RS metric which is an $AdS_5$ metric must be reduced from 5+2 dimensions, in view of Refs.\cite{4bars,5bars}. This is an interesting issue that needs to be studied in a way that is compatible with the views in Refs.\cite{4bars,5bars}. We also see that there are several studies on Dilaton in cosmological problems such as the cosmological constant and inflation \cite{42a,42b}. However, inflation scenarios and reheating periods have not been enough considered in this model. We have only initially estimated that the Dilaton is inflaton in the slow-roll scenario. This shows another opportunity to study inflation as well as extra dimensions in the chaotic and hybrid inflation framework.

The last, the model can be viewed as a hologram of string theory and a connection between string theory and the SM. So studying and extending this model is an interesting way to create a good bridge between string theory and the SM, in which the problem of inflation and baryogenesis are two very significant constraints.

\section*{ACKNOWLEDGMENTS}

In memory of the day our venerable teacher, Dr. Vo Thanh Van, passed away on 5 July 2016, this series of our papers highlights the continuation of what he has done through many years in the Department of Theoretical Physics. This research is funded by Vietnam National Foundation for Science and Technology Development (Nafosted) under grant number 103.01-2021.09.
\\[0.3cm]

\end{document}